\documentclass[conference,compsoc]{IEEEtran}
\ifCLASSOPTIONcompsoc
  \usepackage[nocompress]{cite}
\else
  \usepackage{cite}
\fi
\usepackage{url}

\ifCLASSINFOpdf
   \usepackage[pdftex]{graphicx}
   \usepackage{rotating}
   \usepackage{soul}
   \usepackage[table]{xcolor}
\else
\fi
\begin{document}
\title{A structural study of Big Tech firm-switching of inventors in the post-recession era}

\author{\IEEEauthorblockN{Yidan Sun}
\IEEEauthorblockA{Information Sciences Institute\\
University of Southern California\\
Marina del Rey, California, United States 90292\\
Email: yidans@isi.edu}
\and
\IEEEauthorblockN{Mayank Kejriwal}
\IEEEauthorblockA{Information Sciences Institute\\
University of Southern California\\
Marina del Rey, California, United States 90292\\
Email: kejriwal@isi.edu}}
% \and
% \IEEEauthorblockN{James Kirk\\ and Montgomery Scott}
% \IEEEauthorblockA{Starfleet Academy\\
% San Francisco, California 96678-2391\\
% Telephone: (800) 555--1212\\


% compsoc conferences compared to traditional conferences), use this
%Telephone: (800) 555--1212, Fax: (888) 555--1212}


\maketitle

\begin{abstract}
Complex systems research and network science have recently been used to provide novel insights into economic phenomena such as patenting behavior and innovation in firms. Several studies have found that increased mobility of inventors, manifested through firm switching or \emph{transitioning}, is associated with increased overall productivity. This paper proposes a novel structural study of such transitioning inventors, and the role they play in patent co-authorship networks, in a cohort of highly innovative and economically influential companies such as the five Big Tech firms (Apple, Microsoft, Google, Amazon and Meta) in the post-recession period (2010-2022). We formulate and empirically investigate three research questions using Big Tech patent data. Our results show that transitioning inventors tend to have higher degree centrality than the average Big Tech inventor, and that their removal can lead to greater network fragmentation than would be expected by chance. The rate of transition over the 12-year period of study was found to be highest between 2015-2017, suggesting that the Big Tech innovation ecosystem underwent non-trivial shifts during this time. Finally, transition was associated with higher estimated impact of co-authored patents post-transition. 
\end{abstract}


% \fi
\IEEEpeerreviewmaketitle

\section{Introduction}
Innovation has long been recognized in the economics and social sciences for driving long-term productivity and economic measures of aggregate income, such as the Gross Domestic Product (GDP) \cite{inn1, inn2, inn3, inn6}. For private organizations, especially in a knowledge-based economy \cite{kbe1,kbe2} such as is prevalent in much of the industrialized world, innovative products and services developed and commercialized by the firm can be an important driver\footnote{As with any complex system, it bears noting that the relationship between innovation and valuation is itself a complex one, and not always significant (some instances of negative association for specific sectors and types of innovations may be found in, for example, \cite{neg1}). Both the sector and geography can play a role. Most studies have found support for the claim that for technology and `science-based' firms (to quote the terminology used in \cite{inn5}), there is a positive relationship.} of the firm's valuation \cite{inn5,inn6}. Innovation can also have positive societal benefits in the form of higher corporate social responsibility (CSR) by more innovative forms. For example, in an influential recent work, Mishra analyzed a sample of more than 13,000 US `firm-years' over a 15 year period starting from the early 1990s and found that, mediated by high CSR, more innovative firms ended up achieving significantly higher valuation following a period of innovation \cite{inn7}. 

The positive effects of innovation on valuation are especially strong for technology sectors (e.g., fintech \cite{kabulova2020valuation}), and for early-stage companies such as start-ups \cite{greenberg2013small}. Although not the only way of measuring \emph{innovation productivity} in an organization, the number of patents filed by the organization is an objective measure that has been widely studied by researchers and policymakers alike\cite{sorensen2000,burhan2017,chen2009}. At the time of writing, there is an enormous body of literature on constructing and studying patent networks, estimation of valuation from patent analysis (including network analysis), textual analysis of patents, as well as a range of qualitative studies on patents\cite{yoon2004,yoon2011,patentval,patentqual}. 

% Despite this proliferation of broad patent studies, studies aimed at a sector or a smaller cohort of important companies over a reasonable time period are relatively rare. An example of such a cohort is the set of five `Big Tech' firms (Apple, Google, Meta, Microsoft, and Amazon). In the aftermath of the 2008 financial crisis, these companies have ended up outperforming the broader market by a considerable margin and are among the most valuable companies in the world according to their recent market valuation\footnote{https://www.cnbc.com/2020/01/28/sp-500-dominated-by-apple- microsoft-alphabet-amazon-facebook.html}. 

In this paper, we use patent analysis, including construction and study of a patent co-authorship network, to understand the structural properties of \emph{firm-switching} or \emph{transitioning} inventors. We focus our study on a cohort of five `Big Tech' firms (Apple, Google, Meta, Microsoft, and Amazon) that, in the aftermath of the 2008 financial crisis, have ended up outperforming the broader market by a considerable margin and are among the most valuable companies in the world\footnote{https://www.cnbc.com/2020/01/28/sp-500-dominated-by-apple- microsoft-alphabet-amazon-facebook.html}. Here we define an inventor as co-author on a patent where one of these five organizations serves as the `assignee' organization i.e., the organization that files the patent, and to whom the intellectual property legally  belongs. A transitioning inventor is one who is originally employed by one of the five Big Tech firms (e.g., Apple), and is co-author on patents filed by Apple, but subsequently switches to a different firm (e.g., Amazon) and starts co-authoring patents filed by that firm. 

There is increasing evidence that such firm-switching or `transitioning' inventors may play an under-appreciated role in boosting innovation potential within the technology ecosystem \cite{inventormob1}. Intuitively, the ability to transition represents inventor `mobility' \cite{hoisl2007tracing}, and even for general workers (not just inventors), evidence suggests that mobility can lead to productivity gains and better incentive alignment in the overall economy \cite{mob2}. Highly productive inventors are also more likely to be `poached' by a rival company that is looking to innovate in a similar area. The rival company may be incentivized to offer greater benefits, including higher pay and more freedom to the inventor in developing and publishing their ideas, many of which the organization can potentially look to patent. Because poaching is usually targeted, in theory, this can motivate the inventor further due to greater alignment between the company's incentives and the inventor's goals. While firm switching behavior has received some theoretical and empirical attention in the economics literature \cite{inventormob1, mob2, mob3}, it has never before been studied from a structural perspective (i.e., from the perspective of \emph{economic complexity} \cite{econcomplex}), in the five Big Tech organizations in the post-recessionary era (2010-2022).

% This paper proposes a novel structural study of transitioning inventors in the five Big Tech firms listed above during the post-recessionary period of 2010-2022. 
We formulate and investigate three specific research questions (RQs) to better understand the structural role and other properties of these transitioning inventors in the broader Big Tech ecosystem during this period:

%Are transitioning inventors more central%
{\bf RQ1:}  How does the degree distribution and other structural properties of the transitioning inventors in the patent co-authorship network compared with those of the other inventors, and does the removal of these inventors from the network lead to lower connectivity and more fragmentation than would be expected through chance? 

% How does the removal of transitioning inventors impact the structural properties and connectivity patterns of the co-author patent network?}

{\bf RQ2:} Does the rate of inventor transition remain constant (or a show a monotonic trend) over time, or is there a specific period during which the majority of transitions occur?

{\bf RQ3:} Is transition associated with an increase in estimated impact of co-authored patents (reflected through a subsequently defined measure, such as the normalized number of citations received by their patents) \emph{after} transition, as compared to the impact of the patents co-authored \emph{before} transition?

The first research question explores the structural properties of transitioning inventors using the usual tools of network science. We consider these properties both with respect to the overall co-authorship network (that includes all patent authors within Big Tech in the period under study, including the vast majority of patent authors in Big Tech who have not transitioned) and also the `sub-network' where only the transitioning inventors are represented as nodes. The second research question is instead considering what happens to the rate of transition over the period under study; do we, for example, see relatively stable and constant rates, steadily increasing or decreasing rates, or a trend that is more complex? We also consider whether some organizations experience greater transitions (whether incoming or outgoing) than others or if the transition-mixture is relatively even among the five Big Tech firms. Finally, the motivation behind studying the third research question is to understand whether transition is associated with an estimated measure of `excess' impact due to transition. While we only consider one measure of excess impact, our definition attempts to control for potential biases, such as the length of time elapsed since the patent was granted.

\section{Materials and Methods}

% \subsection{Data}
Our research primarily focuses on analyzing patents filed by five major technology corporations: Google, Meta (formerly known as Facebook), Apple, Amazon, and Microsoft. To gather the relevant patent data, we utilized the Query Builder feature in the PatentsView platform \cite{patentsview}. PatentsView is a comprehensive visualization, data dissemination, and analysis tool specifically designed for intellectual property (IP) data. The platform receives support from the Office of the Chief Economist at the U.S. Patent \& Trademark Office (USPTO). The Query Builder feature within PatentsView allows researchers to identify and retrieve specific subsets of patents of interest from the vast collection of all U.S. patents. To obtain patent data for this study, we issued the following query to Query Builder: `\emph{Assignee Organization} contains\footnote{The special query keyword `contains'  will return any patent with an assignee organization that has the specified organization name as a substring. For example, a patent filed by `Google LLC' as the assignee organization will be returned if the organization name is specified as `Google'. While we could use the special keyword `equals' rather than contains, it is less robust than the former and does not capture subsidiary organizations, as discussed subsequently.} \emph{[Organization Name]}'. Here, the \emph{[Organization Name]} is a placeholder for each of the five Big Tech companies we targeted and the \emph{Assignee Organization} is a field in the PatentsView dataset. We also specified a time constraint as we wanted to focus on patents filed by these companies between 2010 and 2022 (inclusive), and their patents granted within this period.

During the query process, we noticed that there were instances where other irrelevant assignees included similar keywords. For example, when searching for `\emph{Assignee Organization} contains \emph{Apple}', entries such as `Appleton Papers Inc' would also be retrieved, which is unrelated to the company that we intended to study. To maintain the accuracy of our analysis, we cross-verified the patents and inventors associated with each retrieved assignee, using external sources, by identifying the assignees that are genuinely related to the companies we intended to study.

Furthermore, these technology corporations often have numerous branches, alternate names, and variations in spelling. For instance, Google may be referred to as Google LLC, Google Technology Holdings LLC, or even encounter misspellings like Google LCC. To ensure a comprehensive analysis, we consolidated all subsidiary branches and alternate names under one unified name. In this example, we merged all variations into `Google'. The same approach is also applied to the other four companies being studied, including Meta, Apple, Amazon, and Microsoft.

The complete set of fields\footnote{A data dictionary describing all of these fields may be accessed at \url{https://patentsview.org/query/data-dictionary}.} available for each patent in the PatentsView Query Builder can be quite extensive (including the patent text). Our analysis is restricted to a selected subset of these fields. These fields of interest include:
\begin{itemize}
    \item \texttt{app\_date}: The date when the patent application was filed.
    \item \texttt{assignee\_organization}: The organization or company to which the patent rights have been assigned. In our study, these organizations refer to the five companies under investigation: Google, Meta, Apple, Amazon, and Microsoft. Each patent can only have one assignee organization.
    \item \texttt{inventor\_key\_id}: A unique identifier for each inventor.
    \item \texttt{inventor\_first\_name}: The first name of the inventor(s) listed on the patent.
    \item \texttt{inventor\_last\_name}: The last name of the inventor(s) listed on the patent.
    \item \texttt{patent\_title}: The title or name given to the invention covered by the patent.
    \item \texttt{patent\_date}: The date when the patent was granted.
    \item \texttt{patent\_number}: The unique identification number assigned to the patent.
    \item \texttt{citedby\_patent\_number}: The number of other patents that have cited this particular patent.
\end{itemize}

\begin{figure*}
\centering
\includegraphics[width=7in]{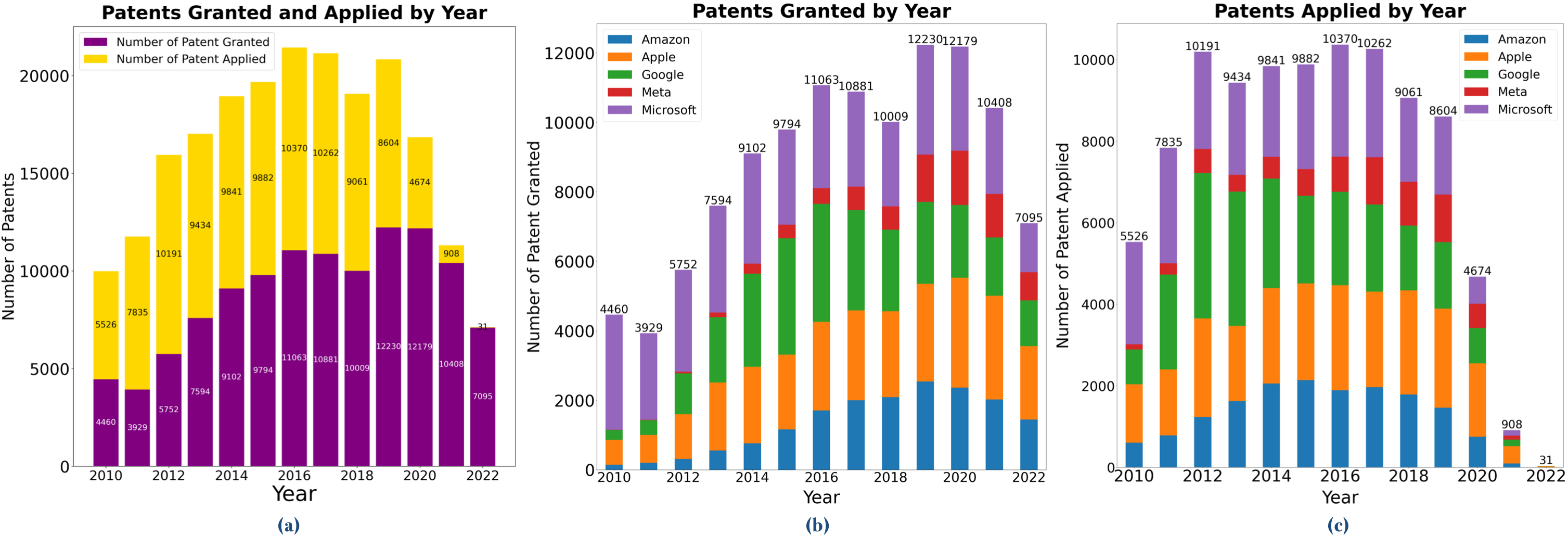}
\caption{(a) The distribution of granted patents and filed patents; (b) distribution of patents granted per year, by company; (c) distribution of patents applied for by each company per year.}\label{fig1}
\end{figure*}

% We focused our analysis on specific columns that provide necessary information for our study: application dates (\texttt{app\_date}), inventors, the changes in their \texttt{assignee\_organization}, and the citations each patent receives (\texttt{citedby\_patent\_number}). 
% After retrieving the data, we found that, since 1976, a total of 79,209 inventors have filed patents associated with the five companies, or assignee organizations, under investigation.

{\bf Basic descriptive statistics.} After retrieving the data, we found that, since 2010 (and up until 2022), 2,329 individuals have been granted patents with multiple assignee organizations. More details on the patents filed and granted, distributed across the five organizations, are illustrated in Figure \ref{fig1}. Among the 2,329 inventors with multiple assignee organizations, 2,213 inventors have patents associated with exactly two different assignee organizations. In relative terms, the number of individuals in these five companies (between 2010 and 2022) authoring patents with multiple assignee organizations is a small, but non-trivial, fraction of the total number of individuals authoring patents with these five organizations (74,637 inventors).

We ordered the patents submitted by each of these 2,213 inventors according to the application date of the patent. Within this group, 873 inventors alternated between two different assignee organizations multiple times, filing patents under one assignee and then switching to the other, and vice versa. On the other hand, 1,340 of these inventors transitioned only once, initially filing patents for one assignee organization and subsequently moving to another.

{\bf Qualitative spot-checks.} To gain preliminary qualitative insights into these 1,340 inventors' professional trajectories, we randomly selected a subset of 30 inventors from the group of 1,340 and conducted a preliminary investigation of their profiles using external sources such as LinkedIn. Our findings revealed that all 30 inventors had undergone transitions from one company to another among the five companies under investigation throughout their careers.  Importantly, we observed a correlation between the timing of these job transitions and the periods when they started filing patents with their subsequent company, according to their patent application dates (\texttt{app\_date}). Furthermore, the companies they were employed by corresponded with the companies they assigned their patents to. This suggests that a change in the patent assignee often indicates a job transition, rather than holding concurrent positions at two different companies. Therefore, in this study, we assumed that for the 1,340 inventors who changed their patent assignees once, this shift likely represented a move from one company to another among the five companies being investigated.

{\bf Co-inventor patent network (CPN) construction.} We constructed the co-inventor patent network (CPN), denoted here as $C_1=(V,E)$, by using authorship data from all of the patents collected ($|V|=74,637$), as described in the previous section. In other words, each unique inventor was assigned a unique node in the CPN, and an (undirected, unweighted) edge was created between two inventors if they co-authored a patent together as employees in a Big Tech firm during the period of interest (2010-2022). 

Recall that the subset of these inventors who transitioned just once between different organizations from 2010 to 2022 consists of 1,340 inventors. To understand the structural properties of these inventors with respect to the overall co-inventor network, we also constructed the subgraph $C_2 \subset C_1$, of which the nodes consists of these 1,340 inventors, and the edges are the subset of edges in $C_1$ that can only exist between these nodes. The goal behind constructing $C_2$ is to study the unique collaboration dynamics of inventors who have experienced a transition in their organizational affiliation. Finally, for investigating the fragmentation aspect of RQ1, we also consider the network $C_r$, which is obtained by removing all nodes in $C_2$ from $C_1$ (in other words, the 1,340 transitioning inventors), and all edges in $C_1$ that were incident on at least one of the 1,340 inventors.

% To understand the impact of these transitioning inventors on the overall collaboration network, we created another network $C_r$ by removing the nodes in $C_2$ from $C_1$. $C_r$ represents the remaining structure of the co-author patent network after the transitioning inventors have been removed. 

\section{Results and Discussion}

\subsection{RQ1}
\begin{figure*}
\centering
\includegraphics[width=7.0in]{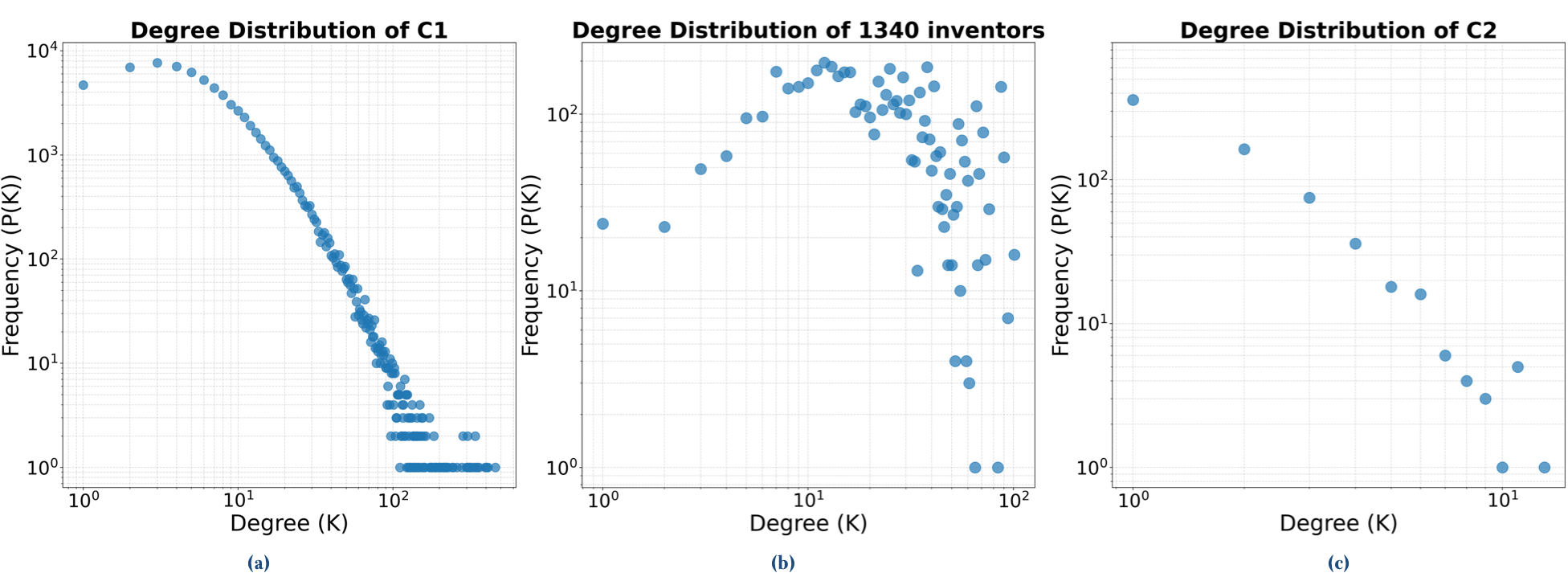}
\caption{(a) The degree distribution of all nodes in the CPN $C_1$; (b) degree distribution of the subset of 1,340 transitioning inventor nodes in $C_1$; (c) degree distribution of all nodes in $C_2$.}\label{fig2}
\end{figure*}
The original CPN ($C_1$) consisting of 74,637 inventors (as nodes) is connected by 372,245 edges or unique co-authorships\footnote{Because the CPN is undirected, a co-authorship relation is counted only once. In the future, we will also consider the weighted version of this network.}. The degree distribution of $C_1$ follows a power-law distribution with a gamma value of -2.32. Despite the large number of inventors, only a small fraction ($\sim$2.5\% or 1,900 nodes) are singletons, indicating a relatively well-connected network. The average inventor in this network has co-authored patents with approximately 10 other inventors, with the most prolific inventor having co-authored with 460 others. The network is divided into 3,598 connected components (including singletons), with the largest connected component consisting of 66,818 inventors. The density of the network is relatively low ($\sim$0.000134), implying that it is still quite sparse, despite the inventors being divided only among five unique organizations. However, as expected from the social nature of such collaborations, the average clustering coefficient is moderately high ($\sim$0.675), which indicates substantial clustering and transitivity among the inventors. 

We also computed the degree distribution of the nodes corresponding to the 1,340 inventors who transitioned only once from 2010 to 2022 within the network graph $C_1$. Figure 2 (a) and (b) illustrate the overall degree distribution (all nodes) in $C_1$ as well as the degree distribution of the subset of 1,340 nodes in $C_1$. We then ranked these degrees and compared the ranks with respect to the overall network. We found that 80.61\% of the transitioning inventors are in the top quartile (25\%) of all nodes, and 95.07\% of the transitioning inventors are within the top half (50\%) of all nodes. These are significantly higher than would be expected for a random sample of 1,340 nodes selected from the network. These numbers suggest that the majority of transitioning inventors have high degree centrality, indicating some degree of influence (at least in a local sense) within the network. However, by calculating the clustering coefficient for each of the transitioning inventors in $C_1$ and subsequently averaging these coefficients, we derive an average local clustering coefficient of approximately 0.379 for the these transitioning inventors, which is lower than the network average of 0.675. This suggests that transitioning inventors are structurally embedded in a more star-like structure: all collaborators of transitioning inventors do not themselves collaborate on the same patents as often. In other words, transitioning inventors are co-authoring patents with groups of inventors between whom the only connection is the transitioning inventor themselves. 

Recall that we had also constructed the subgraph $C_2$ that only contains as nodes the 1,340 inventors who transitioned just once from 2010 to 2022. We found that $C_2$ is denser ($\sim$0.000791) than $C_1$ while having lower average clustering coefficient (0.0986) than $C_1$. Among the 1,340 inventors, there are 710 co-authorship connections, but nearly half of the 1,340 inventors have not co-authored patents with any others in $C_2$ (although they have co-authored patents with inventors in the larger network $C_1$). These transitioning inventors therefore have fewer co-authorship connections on average, as evidenced by an average degree of approximately 1.06 and a maximum degree of 13. The shape of this degree distribution is illustrated in Figure 2 (c). While the data is too small to conclusively interpret the shape as that of a scale-free distribution, it is also not inconsistent with such a distribution. 

The `remaining' network, $C_r$, which mainly consists of inventors who have not transitioned, has a slightly lower average number of co-authorship connections per inventor compared to $C_1$. Additionally, the number of inventors without any co-authorship increases, suggesting that the removal of transitioning inventors led to some inventors in $C_1$ becoming isolated. We explore fragmentation due to the removal of these nodes in more detail subsequently. As expected, the density of $C_r$ is lower than that of $C_1$, indicating less intense collaboration among the remaining inventors.

% Note[The below result is ok but not strong enough to me; I just came across that I should try sampling non-transitioning inventors (i.e. those have never changed company), to see whether that will improve; will do the experiment tmr] 

We evaluated the fragmentation caused by the removal of these 1,340 inventors on the CPN by comparing the number of connected components and the average clustering coefficient between the original network ($C_1$) and $C_r$ (the network after removing the transitioning inventors), as well as a network generated by randomly removing 1,340 inventors from the original network. Compared to $C_1$, the average number of connected components in $C_r$ increases to 3,952. This indicates that transitioning inventors play a bridging role in the innovation ecosystem, which was also suggested earlier by their lower clustering coefficient (but higher degree centrality) compared with the overall network. There are now (i.e., after removing the transitioning inventors) more isolated groups of inventors who are interconnected within their groups but not connected to other groups in the network. 

Note that this fragmentation is greater than would be expected by chance, lending an affirmative answer to the second part of RQ1. To quantify this, we performed an experiment where we \emph{randomly} removed an equivalent number of inventors (1,340) from the network $C_1$, as noted above. We repeated this process independently 100 times, and averaged the number of connected components across these 100 iterations. This average (3,897.1 connected components) is found to be significantly lower than 3,952 (N=100, p = $1.46 \times 10^{-9}$, one-sided Student's t-test). This suggests that transitioning inventors play a greater role in maintaining the network's global connectivity than a random group of inventors embedded in the same environment (the five Big Tech firms), as their removal results in slightly increased fragmentation compared to the removal of a random set of inventors.

Similarly, the average clustering coefficient exhibits interesting behavior upon the removal of inventors. The average clustering coefficient in the remaining network ($C_r$) is 0.674. This suggests that, on average, the remaining inventors in $C_r$ still tend to form moderately interconnected groups. Despite the network fragmentation caused by the removal of the 1,340 inventors, collaboration among groups of inventors remains largely unaffected, as indicated by the similar average clustering coefficient in $C_r$ and $C_1$.

In contrast, when we repeatedly remove a random set of 1,340 inventors  from the network, the average clustering coefficient of the nodes in the remaining network is slightly lower, at 0.660. This value is also found to be significantly lower than 0.674 (N=100, p = $4.074 \times 10^{-20}$, one-sided Student's t-test). This indicates that the removal of the specific group of transitioning inventors results in a network ($C_r$) that maintains slightly higher transitivity as compared to the network resulting from the random removal of inventors.

\subsection{RQ2} 
The primary focus of RQ2 is to examine if the rate of transition of the 1,340 inventors from 2010 to 2022 is relatively stable over the period of study, or if it peaks (or shows other irregularities) during some periods. We also seek to investigate if transitioning occurred uniformly often between the five organizations, or if some organizations are, in fact, over-represented in terms of out-transitions or in-transitions than would be expected through chance alone. As a first step toward these investigations, we determined an estimator for the transition time $T_t$ when an inventor transitioned from the first organization to the second organization. To do so, we identified when a transitioning inventor filed their \emph{last} patent with the first assignee organization and when they filed their \emph{first} patent with the second assignee organization. The estimate of $T_t$ was then calculated the midpoint between these two points in time\footnote{In a slight abuse of terminology, we continue to refer to this estimator as $T_t$, rather than the unknown variable itself (mathematically representing a more `exact' time when the inventor transitioned from the first to the second organization) that is being estimated.} 
\begin{figure}
\centering
\includegraphics[width=1\linewidth]{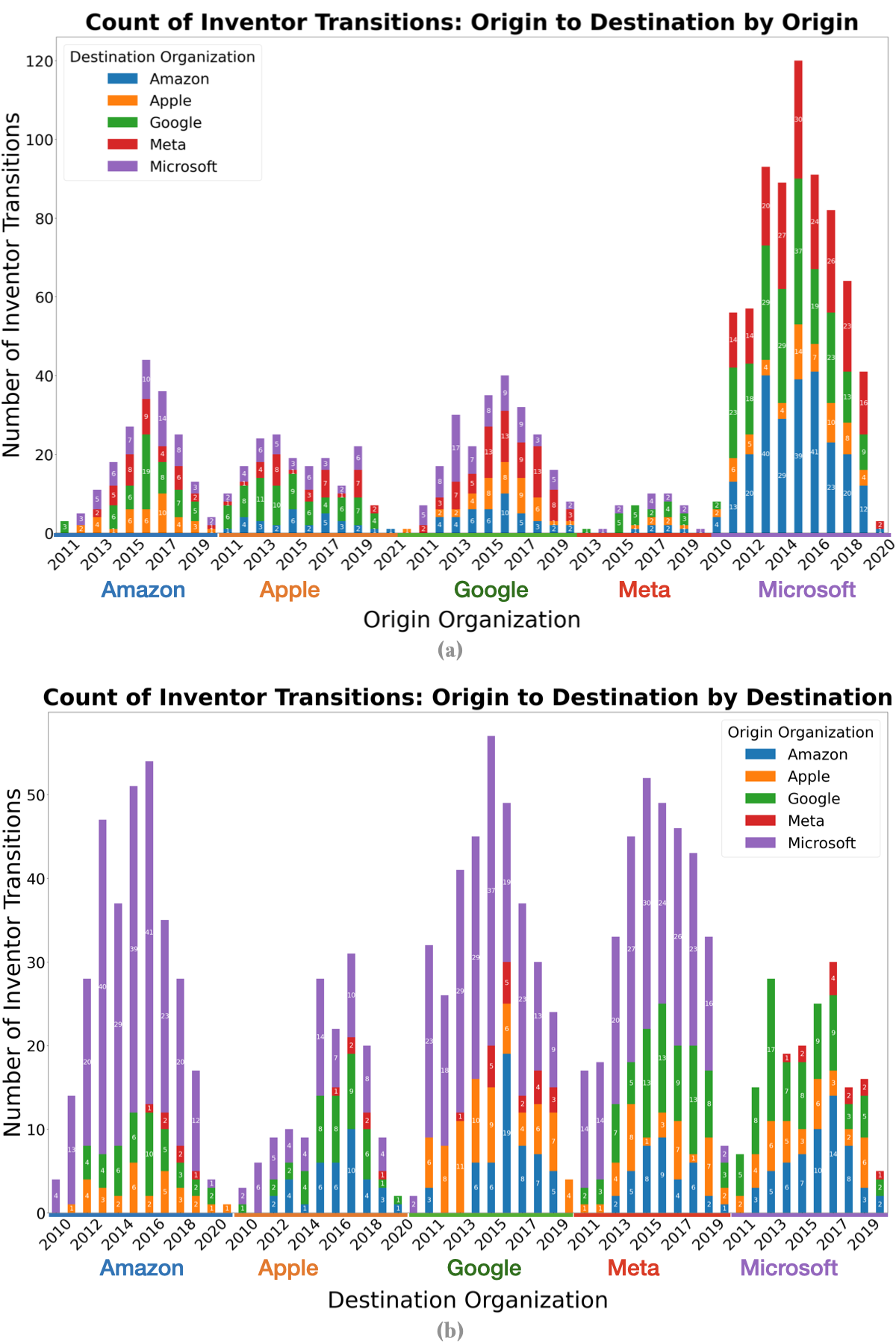}
\caption{The number of inventor-transitions from their first assignee organization to their second assignee organization each year from 2010 to 2022. (a) The number of inventor-transitions outflow by company, with the color in each bar indicating the destination company the inventor transitioned to. (b) The number of inventor-transitions inflow by company, with the color in each bar indicating the source company the inventor transitioned from.}\label{figRQ2:1}
\end{figure}

Figure \ref{figRQ2:1} illustrates the number of inventors transitioning from their initial assignee organization to their second assignee organization each year. We show company-segregated distributions for both incoming and outgoing transitions (and for each year) to gain deeper insights into the trend. Among the 1,340 inventors who transitioned just once from 2010 to 2022, Microsoft, on average, had the most significant outflow of inventors, while Meta had the lowest. Regarding transitions into the companies, Google, Amazon, and Meta have the highest average influx of inventors. Therefore, outflowing transitions were significantly more skewed than incoming transitions, which were more evenly divided. Furthermore, the annual rate of inventor transitions suggests that there is a peak from 2015 to 2017 for both outgoing and incoming transitions. Together, the figure shows a concentrated transitioning period between the assignee organizations, which further implies that the innovation ecosystem within these organizations underwent a non-trivial shift (with the greatest changes suggested for Microsoft) during a relatively brief 3-year period. 

While the results in Figure \ref{figRQ2:1} are illustrative, they are tabulated at the granularity of years. To gain further insight into these transitions at a finer granularity (which is also useful because the rate of transition increases during a relatively narrow time period of 2-3 years), we compute two additional variables for each transitioning inventor: first, we record the time when they first filed a patent with their first assignee organization as their `innovation start' time $T_s$. Similarly, we marked the time when they last filed a patent with their second assignee organization as their `innovation end' time $T_e$. This period between the start and end times is referred to as the `innovation period'\footnote{It bears noting that these terms should only be interpreted in the context of this study and time period e.g., some inventors may have been innovating in other organizations (e.g., startups acquired by some of these organizations, or research labs) before or after the time period of study. The `innovation end' time therefore should not be `literally' taken to imply that the inventor has stopped innovating or filing patents.}. Given these additional variables, we sought to identify the periods during which transitions occurred most frequently at a finer granularity (quarters or 3-month periods, rather than years). To achieve this, we utilized a moving three-month window that starts from January 2010 and is incrementally rolled (in increments of one day) all the way until December 2022. Within each such window, we determined how \emph{long} each inventor was associated with either the first or second assignee.

\begin{figure}
\centering
\includegraphics[width=1\linewidth]{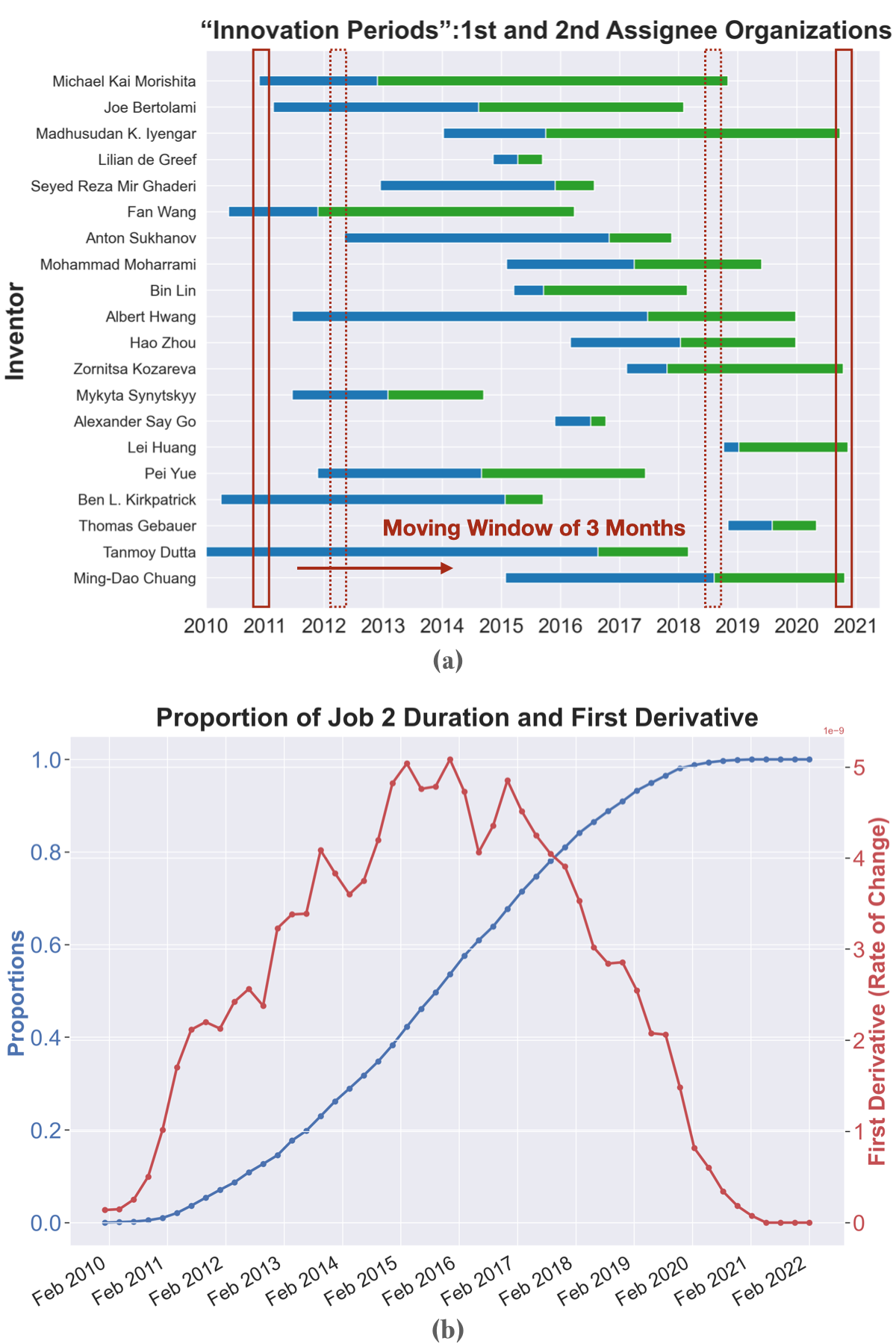}
\caption{(a) An illustration of the 3-month moving window for a set of 20 inventors. The start of the blue bar ($T_s$) marks the beginning of the inventor's `innovation start' time. The transition time to the second organization is denoted by $T_t$, where the blue bar ends and the green bar begins. The end of the green bar ($T_e$) marks the `innovation end' time; (b) The trend of the ratio $\frac{D2}{D1}$ (`proportion of the job 2 duration') between 2010 and 2022, and its first derivative.}\label{figRQ2:2}
\end{figure}

Formally, this calculation may be described as follows: first, for every inventor $i$ in each window, we calculated the length of time spent innovating in the first assignee organization. We do so by measuring the time from the window's start (or from $T_s$ if it commenced after the window began) to either the window's end or $T_t$ (the time of transition), whichever was sooner. This quantity represents the duration that $i$ filed patents with, when they were employed by their first assignee organization. We denote this duration as $D^1_i$. Similarly, we can compute such a duration ($D^2_i$) where $i$ filed patents with their second assignee organization, by measuring the length of time from the later of the window's start, or $T_t$,  and ending at the earlier of the window's end or $T_e$. An example of this procedure using actual data is illustrated for a small set of transitioning inventors in Figure \ref{figRQ2:2} (a).

 Using this procedure, we calculate the total `duration' of innovation in the first organization within a window, by summing all individual durations (namely., the `blue' bars in Figure \ref{figRQ2:2} (a) within a `red' window, but for all transitioning inventors rather than just the ones shown in the figure) for the first organization. We denote this total duration as $D1$. Similarly, we can compute the total duration for the second job (the sum of all the green bars in a red window) for all inventors within a window, as $D2$. Denoting the number of all transitioning inventors as $I = 1340$, we can now compute the proportion $\frac{D2}{D1}$ within each window, and plot this proportion against the mid-point of the window (which spans 3 months). 

The results are illustrated in Figure \ref{figRQ2:2} (b). We can use the plot to determine when transition, as a measure of this relative proportion of innovation activity in the first versus the second assignee organization, exhibits the most significant change. We do so by examining the slope of this curve. We are specifically interested in identifying a two-year span with the steepest slope. A steeper slope indicates a more rapid change in the ratio $\frac{D2}{D1}$. Our analysis indicated that the derivative of the segment from December 2014 to November 2016 remains high (with a short blip in late summer 2016) before declining secularly afterward. This supports the previous result that the most pronounced transition happened in this period, but it also shows that rates had been increasing leading up to this period, before declining continuously after this period. Further investigation of this period, as well as hypothetical causes for a sustained high transition rate, could be an interesting area for further sociological research exploring innovation patterns in Big Tech. 

% . The total duration of the initial jobs, $D1$, and the total duration of the subsequent jobs, $D2$, can be represented mathematically as $D1 = \sum_{i=1}^{|I|} D1_i$, and $D2 = \sum_{i=1}^{|I|} D2_i$. $D1_i$ represents the duration of the initial job for inventor $i$ and $D2_i$ represents the duration of the subsequent job for inventor $i$. To determine the proportion between $D2$ and $D1$ within each window, we calculated the ratio as $\frac{D2}{D1}$. By computing this ratio for each window and plotting the resulting proportions, we obtained Figure 3 (b).

% Finding the period in which the curve exhibits the most significant change can be achieved by examining the slopes of the curve. We are specifically interested in identifying a two-year span with the steepest slope. A steeper slope indicates a more rapid change in the ratio $\frac{D2}{D1}$. Our analysis indicated that the curve segment from December 2015 to December 2017 has the steepest ascent, indicating the most pronounced change in the proportion of the second job. This finding aligns with our previous observation.

\subsection{RQ3} 

RQ3 aims to investigate whether the transition between assignee organizations has had a positive impact on the inventors' ability to produce more high-impact inventions. We note at the outset that determining impact of an innovation, especially in the technology industry, is a controversial topic \cite{jaffe2017patent}, because the impact could occur over long time horizons, be an indirect enabler of other technologies, and is not always measurable using simple metrics. Nevertheless, one way of examining the impact is by considering the citation count of an inventor's patents before and after an inventor transitions. Done in aggregate across all patents filed by the 1,340 transitioning inventors from 2010 through 2022, this count allows us to gauge (in a limited, but still reasonable, manner) the potential association between transition and innovation impact.

One bias that must be accounted for before computing such an association is that patents filed earlier will obviously be expected to receive higher citation counts. Hence, we standardize or `normalize' the citation count $P_{t,o}$ for a patent filed in year $t$ and by an inventor currently in organization $o$ by adopting the following methodology: for each unique combination of year ($t$) and organization ($o$), we calculate the mean ($\bar{P}_{t,o}$) and standard deviation ($s_{t,o}$) of citation counts for all the patents filed within that timeframe and in that organization. We can then normalize $P_{t_o}$ and represent it as a new variable $Z_{t,o}$ using standard Z-score normalization: 
\begin{equation}
Z_{t,o} = \frac{{P_{t,o} - \bar{P}_{t,o}}}{s_{t,o}}
\end{equation}
% \begin{itemize}
% \item $Z$: z-score
% \item $P$: individual patent citation
% \item $M$: mean citation of the company-year combination
% \item $S$: standard deviation of citation
% \end{itemize}

\begin{figure}
\centering
\includegraphics[width=1\linewidth]{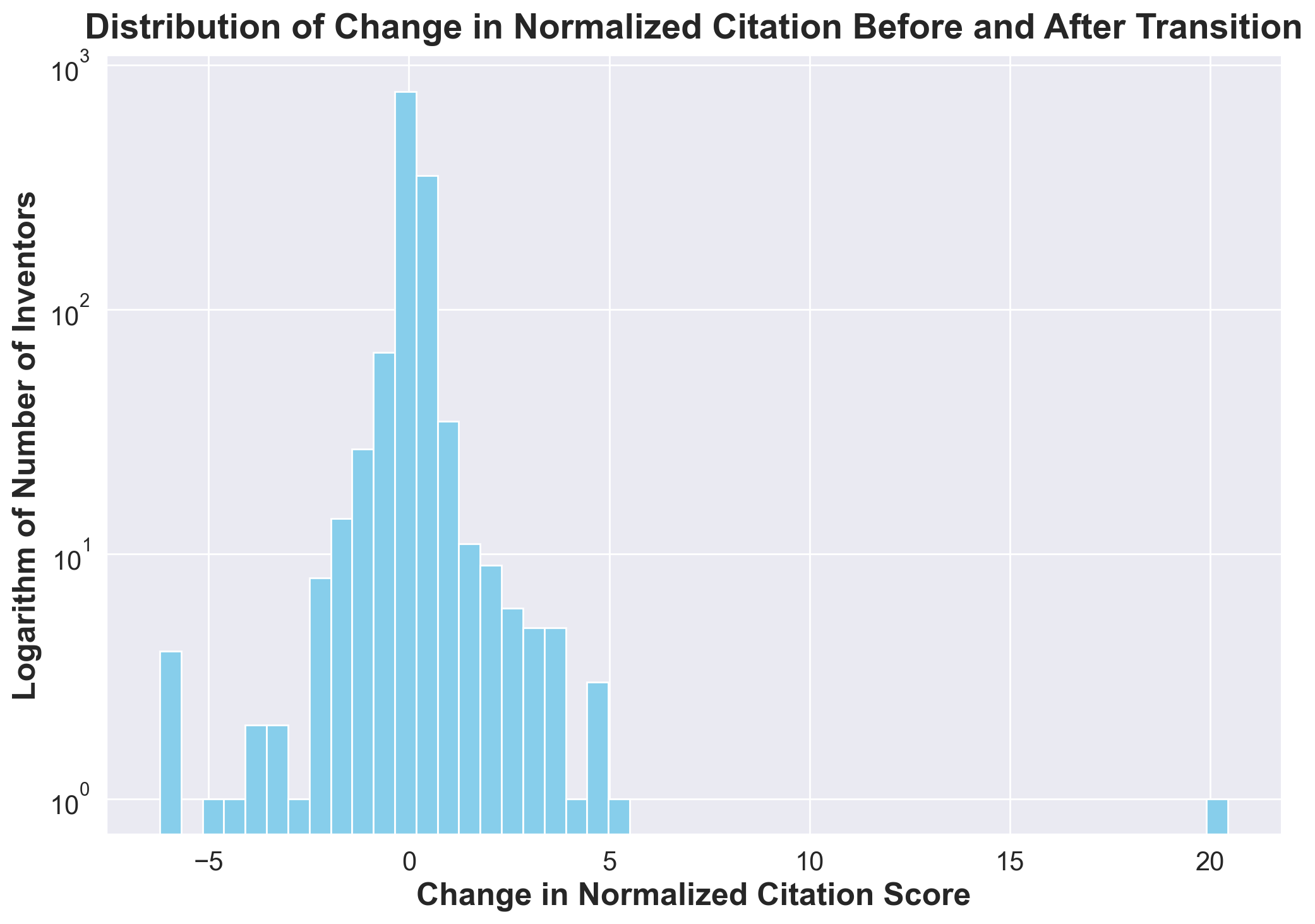}
\caption{Distribution of estimated excess impact, using the methodology described in the main text, for the 1,340 transitioning inventors. Note that the y-axis is on the log scale.}\label{figRQ3}
\end{figure}

By implementing such a Z-score normalization, we are now able to derive a normalized citation count, or estimated impact, for each patent within our dataset. Similarly, we can calculate the estimated impact for an \emph{inventor} over a given time period by averaging the estimated impacts of all patents filed in that time period where that inventor was a co-author. With these measures of estimated impact, we investigated RQ3 by first calculating the average estimated impact for each inventor using the patents filed in the entire time period \emph{before} the inventor transitioned\footnote{Using the terminology introduced earlier for RQ2, this would be the average estimates impact of all patents where the inventor was a co-author in the period $T_t-T_s$; recall that both of these estimators apply individually to each inventor.}, as well as the inventor's average estimated impact \emph{after} the transition. Because these two estimates are obtained for each of the 1,340 transitioning inventors, we obtain two `paired lists' (representing before and after impacts for the same inventor). Subtracting the `before transition' estimated impact from the `after transition' impact, we obtain a single vector containing the estimated \emph{excess} impact (due to transition) for each of the 1,340 inventors. We plot the distribution of this estimated excess impact as a histogram in Figure \ref{figRQ3}. A paired Student's t-test showed evidence in favor of the research hypothesis that the after-transition impact is greater than the before-transition impact (i.e., we were able to reject the null hypothesis convincingly; N=1,340; p= 0.00025).

% The test showed that the null hypothesis can be convincingly rejected (N=1,340; p= 0.00025). In other words, assuming our previous definition and normalization of estimated impact, the test shows that estimated impact after transition  is is (on average) significantly higher than estimated impact before transition. 
While this result would need to be replicated for other measures of estimated impact, and other normalization procedures, it does suggest that the innovation potential of Big Tech has likely been improved, rather than diminished, due to such switching activity (whether voluntarily intended\footnote{Transitioning or switching does not always imply volition on the worker's part. In some cases, the first assignee organization may have terminated the worker, while in other cases, the worker may have been induced to leave and join the second organization. Controlling for volition in such analyses is an important, but difficult, study that future sociological research may want to consider.} by the worker or not).

\section{Conclusion and Future Work}

In investigating RQ1, we found that transitioning inventors can play key structural roles in patent co-authorship networks in Big Tech firms in the period of study, much greater than would be predicted by chance. For example, their removal leads to greater fragmentation, as measured by the change in the number of connected components, compared to the chance removal of an equivalent number of nodes. Results for RQ2 showed that transitioning over this period did not exhibit a regular or `secular' trend: rather, there was a significant outflow of inventors from Microsoft and a prominent inflow into Google, Amazon, and Meta. The rate of transition peaked between 2015 and 2017 and stayed high consistently during this period, suggesting a marked shift in the innovation ecosystem of these organizations during this period. Following this period, the rate of transition declined and ultimately reached a low level by the time of the pandemic. Finally, in investigating RQ3, we found that estimated excess impact, as measured using a normalized citation score of patents that a transitioning inventor co-authored (before and after transition), was positive and statistically significant. This finding suggests that the productivity of transitioning innovators, on average, across Big Tech, was not diminished by transition.

One interesting area of future research is to investigate some of the same research questions but through a causal lens, using additional data (e.g., survey and employment records) and causal inference techniques. Investigating questions like RQ3 by considering a wider body of estimated measure impacts is also a fruitful area for further investigation. Finally, using similar techniques to investigate post-pandemic effects on some of these measures (which will likely only manifest a few years from now) is also of interest.

\bibliographystyle{IEEEtran}
\bibliography{reference}


% \end{thebibliography}

\end{document}